\documentclass[%
 aip,
 amsmath,amssymb,
 preprint,%
 reprint,%
]{revtex4-2}
\usepackage{dcolumn} 
\newcolumntype{T}{D{<}{\leftarrow}{-1}} 

\usepackage{graphicx}
\usepackage{dcolumn}
\usepackage{bm}

\usepackage[utf8]{inputenc}
\usepackage[T1]{fontenc}
\usepackage{mathptmx}
\usepackage{etoolbox}
\usepackage{physics}

\usepackage{multirow} 
\usepackage{paralist}
\usepackage{float}
\usepackage{siunitx}
\usepackage{booktabs}
\usepackage{lipsum}
\usepackage{threeparttable}
\usepackage{amsmath} 
\usepackage{afterpage}
\usepackage{placeins}

\usepackage[dvipsnames]{xcolor} 

\usepackage[colorlinks=true,
            linkcolor=blue,
            citecolor=blue,
            urlcolor=blue]{hyperref}

\let\oldAA\AA
\renewcommand{\AA}{\text{\normalfont\oldAA}}

\makeatletter
\def\@email#1#2{%
 \endgroup
 \patchcmd{\titleblock@produce}
  {\frontmatter@RRAPformat}
  {\frontmatter@RRAPformat{\produce@RRAP{*#1\href{mailto:#2}{#2}}}\frontmatter@RRAPformat}
  {}{}
}%
\makeatother

\newcommand{\vc}{\text{\usefont{OML}{cmm}{m}{it}v}}
\usepackage{amsfonts,bbold,mhchem}
\newcommand{\bmr}{{\bm r}} 
\newcommand{\bmR}{{\bm R}} 
\newcommand{\bmK}{{\bm K}} 
\newcommand{\bmS}{{\bm S}} 
\newcommand{\bmj}{{\bm j}} 
\newcommand{\bmJ}{{\bm J}} 
\newcommand{\bmc}{{\bm c}} 
\newcommand{\bmW}{{\bm W}} 
\newcommand{\bmV}{{\bm V}} 
\newcommand{\bmI}{{\bm I}} 
\newcommand{\bmF}{{\bm F}} 
\newcommand{\bmU}{{\bm U}} 
\newcommand{\bmQ}{{\bm Q}} 

\newcommand{\Half}{\frac{1}{2}}
\newcommand{\Ddf}[2]{\frac{\partial^2 #1}{\partial #2^2}}
\newcommand{\Bra}[1]{\langle #1 |}
\newcommand{\Ket}[1]{| #1 \rangle}

\newcommand{\Threejm}[6]{ \left(\begin{array}{ccc} #1 & #3 & #5\\
                                              #2 & #4 & #6
                                \end{array}
                          \right)}

\newcommand{\Red}{\color{red}}

\newcommand{\Eh}{E$_\mathrm{h}$}

\newcommand{\Qs}{\ensuremath{Q_s}}
\newcommand{\Qa}{\ensuremath{Q_a}}
\newcommand{\Qb}{\ensuremath{Q_b}}
\newcommand{\Qx}{\ensuremath{Q_x}}
\newcommand{\Qy}{\ensuremath{Q_y}}
\newcommand{\Qp}{\ensuremath{Q_+}}
\newcommand{\Qm}{\ensuremath{Q_-}}
\newcommand{\Qpm}{\ensuremath{Q_\pm}}

\newcommand{\hQs}{\hat{Q}_s}
\newcommand{\hQx}{\hat{Q}_x}
\newcommand{\hQy}{\hat{Q}_y}
\newcommand{\hQp}{\hat{Q}_+}
\newcommand{\hQm}{\hat{Q}_-}
\newcommand{\hQpm}{\hat{Q}_\pm}

\newcommand{\ns}{\ensuremath{{n_s}}}

\newcommand{\nb}{\ensuremath{{n_b}}}

\newcommand{\np}{n_+}
\newcommand{\nm}{n_-}

\newcommand{\Ps}{\ensuremath{{p_s}}}
\newcommand{\Pa}{\ensuremath{{p_a}}}
\newcommand{\Pb}{\ensuremath{{p_b}}}
\newcommand{\Px}{\ensuremath{{p_x}}}
\newcommand{\Py}{\ensuremath{{p_y}}}
\newcommand{\Pp}{\ensuremath{{p_+}}}
\newcommand{\Pm}{\ensuremath{{p_-}}}
\newcommand{\Pbmax}{\ensuremath{{p_b^\mathrm{max}}}}
\newcommand{\Psmax}{\ensuremath{{p_s^\mathrm{max}}}}

\newcommand{\gs}{g_s}
\newcommand{\ussss}{u_{ssss}}
\newcommand{\tsss}{t_{sss}}

\newcommand{\tsbb}{t_{sbb}}
\newcommand{\ussbb}{u_{ssbb}}
\newcommand{\ubbbb}{u_{bbbb}}
\newcommand{\uxxxx}{u_{xxxx}}
\newcommand{\uxxyy}{u_{xxyy}}

\newcommand{\has}{\hat{a}_s}

\newcommand{\hap}{\hat{a}_+}
\newcommand{\ham}{\hat{a}_-}
\newcommand{\hapd}{\hat{a}_+^\dagger}
\newcommand{\hamd}{\hat{a}_-^\dagger}
\newcommand{\hasd}{\hat{a}_s^\dagger}
\newcommand{\haxd}{\hat{a}_x^\dagger}
\newcommand{\hayd}{\hat{a}_y^\dagger}
\newcommand{\hapmd}{\hat{a}_\pm^\dagger}
\newcommand{\hamp}{\hat{a}_\mp}
\newcommand{\hapm}{\hat{a}_\pm}

\newcommand{\vs}{\vc_s}
\newcommand{\Vb}{\vc_b}
\newcommand{\lb}{l_b}
\newcommand{\LL}{\lambda}
\newcommand{\LLmax}{\lambda^\mathrm{max}}

\newcommand{\ML}{{\mu}}
\newcommand{\MLmax}{{\mu^\mathrm{max}}}

\newcommand{\tlv}{\tilde{l}_2}
\newcommand{\tOmega}{\tilde{\Omega}}
\begin{document}
\preprint{AIP/123-QED}
\title[\ce{CO2-He}]{Rovibrational cross sections for the \ce{CO2}
$00^00$, $01^10$, $02^20$ and Fermi resonance states
$10^00(1)$ and $10^00(2)$ in collisions with He atoms}
\author{Jelena Bo\v{s}kovi\'{c}}
\affiliation{ 
Kapteyn Astronomical Institute, University of Groningen, NL-9747AD Groningen, The Netherlands}
\affiliation{Theoretical Chemistry, IMM, Radboud University of Nijmegen}
\author{Taha Selim}
\affiliation{Theoretical Chemistry, IMM, Radboud University of Nijmegen}
\author{Ad van der Avoird}
\affiliation{Theoretical Chemistry, IMM, Radboud University of Nijmegen}
\author{Floris F. S. van der Tak}%
\affiliation{ 
Kapteyn Astronomical Institute, University of Groningen, NL-9747AD Groningen, The Netherlands
}%
\affiliation{SRON Netherlands Institute for Space Research, The Netherlands}
\author{Gerrit C. Groenenboom}
\affiliation{%
Theoretical Chemistry, IMM, Radboud University of Nijmegen
}%
\email{gerritg@theochem.ru.nl}
\email{boskovic@astro.rug.nl}
\email{f.f.s.van.der.tak@sron.nl}
\date{\today}
\begin{abstract}
We present quantum scattering calculations of \ce{CO2-He} state-to-state integral
cross sections for 25 transition sets, covering pure rotational transitions
within and rovibrational transitions between: the ground state $00^00$, the
bend fundamental $01^10$, the bending overtone $02^20$, and the Fermi doublet
states $10^00(1)$ and $10^00(2)$. Exact coupled-channel (CC) and
coupled-states approximation (CSA) calculations are performed using a new
five-dimensional potential energy surface.
The calculations cover collision energies from 20 to 5220 cm$^{-1}$, sufficient to obtain converged rate coefficients over 100-800 K, and include initial rotational states for each transition up to at least $j=35$.
We present an energy-grid algorithm for rovibrational quantum scattering calculations that minimizes the number of separate calculations required.
The resulting cross sections exhibit a dominant $\Delta j=0$ propensity and parity-dependent patterns.
For the benchmark $00^00,j' \leftarrow 01^10,j$ transition, thermally averaged rate coefficients reproduce available experimental data. A comparison between the CC and CSA cross sections shows that, for vibrational quenching transitions, the CSA cross sections increasingly deviate from the CC results with increasing initial rotational state $j$. 
Calculated integral cross sections cover collision energies and rotational states needed to derive rate coefficients for non local thermal equilibrium radiative transfer modeling of mid-infrared CO$_2$ spectra.
The computationally simpler \ce{CO2-He} system allows the methodology and
approximations to be tested at high collision energies and rotational states,
helping assess their suitability for the more complex and astrochemically
important \ce{CO2-H2} system.
\end{abstract}
\maketitle
\section{\label{sec:introduction}{Introduction}}
Extraterrestrial CO$_2$ is challenging to observe with ground-based
telescopes because its allowed vibrational transitions
occur at infrared wavelengths that are absorbed by Earth's atmosphere
and its rotational transitions are dipole forbidden.
Interstellar \ce{CO2} was first detected by the IRAS satellite
in 1989\cite{hendecourt:89} through its bend fundamental
around 667 cm$^{-1}$, which corresponding to a wavelength of 15 $\mu$m.
Since then, the \ce{CO2} rovibrational signature has been
widely detected in space, in both ice absorption\cite{bergin:05,
oberg:11} and gas-phase emission\cite{dishoeck:96, boonman:03a,
boonman:03b, carr:08, carr:11, grant:23,
arabhavi:24, vlasblom:25, salyk:25} in many star- and
planet-forming regions, such as molecular cloud cores, protostars and
protoplanetary disks, and recently in spiral arms in our galaxy.\cite{qiu:26}

Rovibrational lines provide valuable diagnostics of astrophysical
environments, as their intensities and shapes depend on the local gas
temperature, density, and molecular abundances. These properties are typically
derived through radiative transfer modeling, which is often performed under the
assumption of local thermodynamic equilibrium (LTE). However, in many
environments, where e.g. James Webb Space Telescope Mid-Infrared Instrument
(JWST/MIRI) observes the upper layers of protoplanetary disks, interstellar gas
densities are too low to maintain LTE (non-LTE regime)\cite{bosman:17}. In
this case, the energy level populations deviate from the Boltzmann
distribution, and collisional rate coefficients are required for non-LTE
radiative transfer modeling of molecular spectra and determination of the
physical and chemical conditions\cite{schoeier:05, roueff:13,
tak:20}. 

While databases such as LAMDA \cite{schoeier:05}, BASECOL
\cite{dubernet:24}, and EMAA \cite{emaa:21,faure:25} provide
collisional data for many interstellar species, collisional data for
vibrationally inelastic transitions remain limited in general, and
state-to-state CO$_2$ rovibrational rate coefficients are still missing.
In this work we compute a large dataset of
energy dependent \ce{CO2-He} cross sections. These results are
used in a separate paper\cite{boskovic:26b} to compute state-to-state
temperature dependent collisional rate coefficients that
are applied to radiative transfer modeling of \ce{CO2} spectra under the
conditions of warm (100-800 K) inner envelopes around protostars and the
inner regions of protoplanetary disks where \ce{CO2} mid-infrared
emissions arises\cite{chen:24c}.
Rotationally elastic and inelastic cross sections important for
pressure broadening in \ce{CO2-He} and \ce{CO2-H2} were
presented recently\cite{niraula:26}.

\begin{figure}[!t]
    \centering
    \includegraphics[width=\hsize]{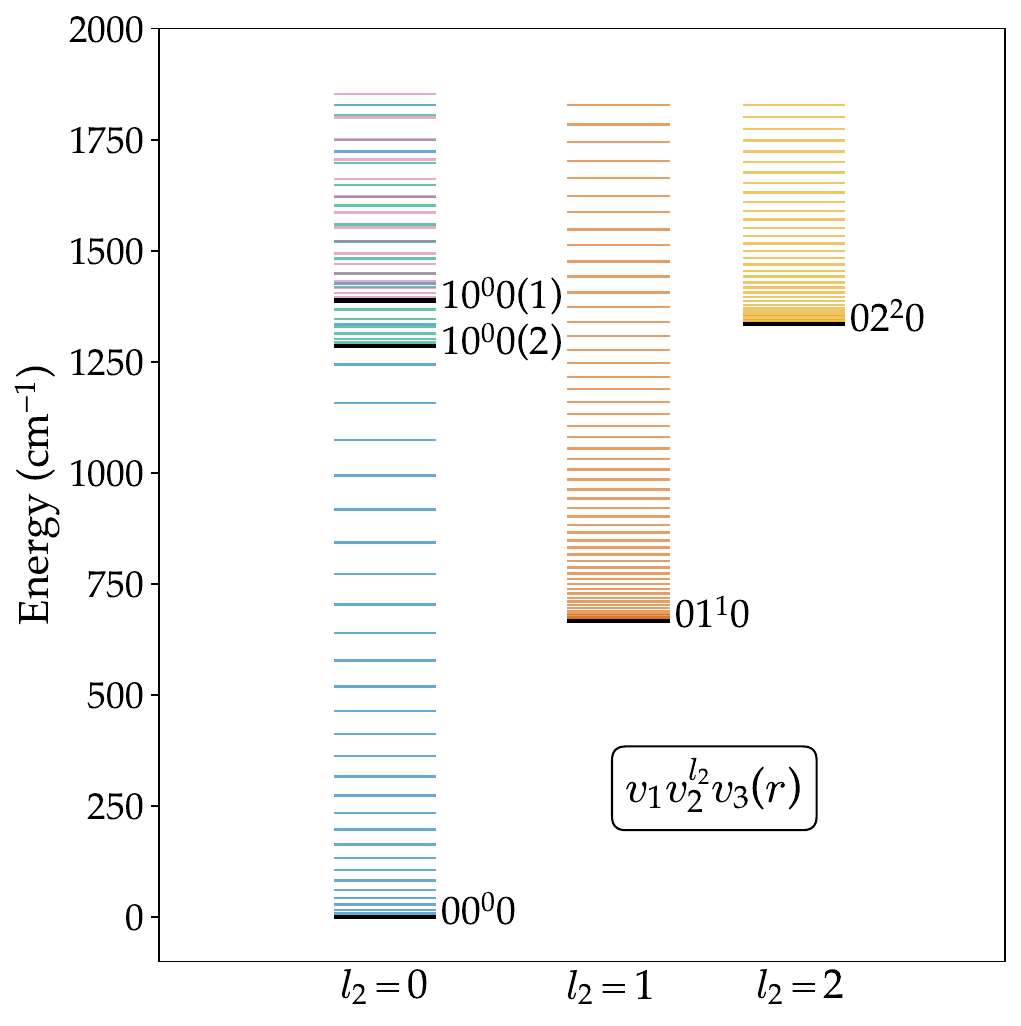}
    \caption{Rovibrational energy diagram of CO$_2$ showing all rovibrational
    states included in our quantum scattering calculations, which are denoted
    as $\vc_1\vc_2^{l_2}\vc_3(r)$, where $\vc_1$ is the symmetric stretch mode,
    $\vc_2$ is the bending mode, $l_2$ is the vibrational angular momentum
    quantum number associated with the bend, $\vc_3$ is the asymmetric stretch
    mode, and $r$ is the resonance number for Fermi resonance states.}
    \label{fig:diag}
\end{figure}

In the 1980s, Clary {\em et al.}\ studied rovibrational transitions
of CO$_2$ in collisions with helium and other rare-gas atoms with
the vibrational coupled channels infinite order sudden (VCC-IOS) method
and with the coupled states approximation (CSA)
\cite{clary:81,clary:81b,clary:82,clary:84,clary:87}.
The present work builds on studies done in Nijmegen on rovibrational transitions between the \ce{CO2} ground state and fundamental symmetric stretch \cite{selim:21,selim:22}, asymmetric stretch \cite{selim:21}, and bend
vibrations \cite{selim:23}. It was established that the VCC-IOS method
fails by orders of magnitude
for ground to symmetric and asymmetric stretch modes \cite{selim:21},
although it gives reasonable quenching rate coefficients for transitions with a smaller energy gap \cite{selim:23}, while the multichannel distorted wave Born
approximation (MC-DWBA) gives good results for the small cross sections
for the stretch modes. Also, the CSA and the more expensive, but also more accurate nearest-neighbor Coriolis coupling (NNCC) method \cite{yang:18} give useful results in some cases, but must always be tested against CC result when used in a new regime.

Here, we expand the study of \ce{CO2-He} collisions in several ways. We compute
a new, five dimensional potential energy surface (PES) which allows us to study
the bend overtone vibrations and their coupling with the symmetric stretch.
With the ``exact'' coupled channels (CC) method we compute cross
sections for all transitions between the rovibrational levels
shown in Fig.~\ref{fig:diag}.
We also compute the pure rotational cross sections for all five
vibrational states, $00^00$, $01^1$0, $02^20$, $10^00(1)$, and $10^00(2)$.
The $\vc_1\vc_2^{l_2}\vc_3(r)$ HITRAN database\cite{gordon:17} notation
of the \ce{CO2} levels is explained in the caption of the figure.
Accurate spectroscopic data on \ce{CO2} can be found
in Ref.~\onlinecite{yurchenko:20}.

Unlike radiative processes, which are governed by strict selection rules,
molecular collisions in principle allow all possible rovibrational
transition, with propensity rules favoring small energy
gaps, $\Delta E = |E_{v'j'}-E_{vj}|$ and small changes
in rotational quantum number, $\Delta j = |j'-j|$ \cite{yang:19c}.
Propensity rules may also depend on the parity of the initial
and final states \cite{alexander:82b}. For rovibrational inelastic
collisions a $\Delta j=0$ propensity was found for atom-diatom
scattering, e.g. for H+CO (Fig.~6 in Ref.~\onlinecite{song:15b}),
and for He+\ce{CO2} (Fig.~7 of Ref.~\onlinecite{clary:83} and 
Fig.~5 of Ref.~\onlinecite{selim:22}). The rule holds in particular for
nonzero initial $j$. In the present work we find that the
$\Delta j=0$ propensity persists for all vibrational transitions
we investigate. We also find clear parity propensity rules, see
Sec.~\ref{sec:results}.

When computing rovibrational cross sections, in general we only 
include the initial and final vibrational states in the channel basis.
However, the stretch fundamental $10^00$ is nearly degenerate with the $l_2=0$
bend overtone $02^00$, so these states mix and give rise to the
$10^00(1)$ and $10^00(2)$ upper and lower Fermi resonance states.
For transitions involving
these states we always include both of them, so the channel basis
may include up to three vibrational states.

We only compute de-excitation cross sections, since the
excitation cross sections follow from microscopic reversibility.
For transitions to lower lying vibrational states we include
all initial rotational levels up to $j=35$, while final rotational
levels can be up to $j'=68$ in the case of transitions from
the upper Fermi level to the vibrational ground state.
Bend fundamental final states may have up to $j'=54$.
We also include transitions to higher lying vibrational levels for
initial states with $j=42,43,\ldots,68$.
These are still de-excitations, since the initial rotational
states are (much) higher than the final rotational states.
The $^{16}$O oxygen atoms in \ce{CO2} are bosons because they
have nuclear spin zero, so only levels with \textit{even}
rotational angular momentum quantum number $j$ exist for vibrational
states with vibrational angular momentum $l_2=0$\cite{selim:23}. 

In previous work on the bend to ground transition\cite{selim:23}
collision energies in the range of 1-2000 cm$^{-1}$ were included.
Below 20 cm$^{-1}$ many resonances were found. Here, we aim to
converge collisional rate coefficients for $T=100-800$ K. In Sec.~\ref{sec:resonances}
we show that the contribution of the resonances is negligible
above 100 K. In order to converge rate coefficients at 800 K we extended
the collision energy to 5220 cm$^{-1}$.

To reduce the cost of the calculations, we made several improvements
to our algorithms and software.
Computing collisional rate coefficients for $T=100-800$ K requires CC calculations for about 30 collision energies, for each initial state.
We devised an algorithm to generate a set of total energies
for which we do CC calculations, such that many calculations
provide cross sections for multiple initial states, reducing
the total number of jobs required by about a factor of ten,
as explained in Sec.\ \ref{sec:Egrid}.

Convergence parameters such as channel basis, number of partial
waves required
and propagation step size depend on the collision
energy, the initial state, and the partial wave.
We do a limited number of test calculations, and interpolate
the required parameters as explained in Sec.~\ref{sec:conv}.

We use the renormalized Numerov method for solving the CC
equations. The original version of B. R. Johnson\cite{johnson:79}
requires two symmetric matrix inversions per step. The implementation by Hubbard {\em et al.} \cite{hubbard:83} is advantageous if the wave function is also needed. However, it involves one matrix multiplication and the solution of a set of linear equations, and comes at the cost of storing two additional matrices. In Appendix \ref{sec:Numerov} we
present an alternative similar in CPU requirements
to Miller's implementation,
but without the additional memory requirement.
The reduced memory requirement is advantageous
when the maximum number of parallel
jobs is limited by available memory. We also reimplemented
the computation of the $\bmK$ and $\bmS$ matrices and the cross sections,
paying attention to memory requirements, since in
a naive implementation in a script language this step,
which contributes little to the CPU time,
becomes the main memory bottleneck.

The CSA method is much cheaper than the CC method, but less
accurate. There have been attempts in the literature to use machine learning (ML)
techniques to correct lower level results for
inelastic collisional rate coefficients\cite{jasinski:20,mihalik:25}.
Thus, we repeated all the calculations with the CSA method
to compare them to CC and we make them available for testing
and developing such ML extrapolations.

As a check of our new potential and our updated software, we
compute temperature dependent vibrational quenching rate coefficients
of the bend fundamental (Sec.\ \ref{sec:resonances}).
 As in a previous work \cite{selim:23},
we find good agreement with the experimental results of
Siddles {\em et al.} \cite{siddles:94}.

Rovibrational inelastic collisions for atom linear-triatom systems
have been studied by Stoecklin {\em et al.} \cite{stoecklin:13,stoecklin:15}.
They used the rigid bender treatment on He+HCN and He-C$_3$.
In the next section we present our method.

\section{Theory}
\label{sec:theory}
\label{sec:qsintro}

We employ CC and CSA calculations with body-fixed (BF)
channel basis functions as
before\cite{selim:22,selim:23}. In this section we
give the main equations and focus on the modifications
and extensions of the method.
In Sec.\ \ref{sec:coords} we define the coordinates, which
now also includes the symmetric stretch, and only use
two-angle embedded frames, which simplifies the presentation somewhat.
In Sec.\ \ref{sec:hamiltonian} we give the Hamiltonian and
the symmetry adapted BF channel basis. The {\em ab initio} calculation
and expansion of the potentials is given in Sec.\ \ref{sec:PES}.
The expansion is similar as before, but we now also include a derivation
of the functional form in Appendix \ref{sec:appendixV}.
The vibrational wave functions of \ce{CO2} are expanded in
a harmonic oscillator basis for bend and symmetric stretch normal
modes, using creation and annihilation operators, allowing us
to compute matrix elements analytically (Secs.\ \ref{sec:normal}
and \ref{sec:matrixelements}). The design of the total
energy grid is discussed in  Sec.\ \ref{sec:Egrid} and the
actual algorithm is given in Appendix \ref{app:Egrid}.

\subsection{Coordinates and frames}
\label{sec:coords}
We define the Jacobi scattering vector as
\mbox{$\bmR \equiv \bmr_\mathrm{He}^\mathrm{SF}-\bmr_0^\mathrm{SF}$},
where $\bmr_\mathrm{He}^\mathrm{SF}$ and $\bmr_0^\mathrm{SF}$
are the space-fixed (SF) Cartesian coordinates of
the helium atom and of the center-of-mass (c.o.m.) of
CO$_2$, respectively. The length of this vector is $R=|\bmR|$.
The positions of the atoms
in the molecule are defined with respect to a reference
geometry, for which we take the linear equilibrium
structure of CO$_2$, with the carbon atom at its c.o.m.
We define $\tilde{\bmr} \equiv \tilde{\bmr}_1^\mathrm{SF}-\tilde{\bmr}_2^\mathrm{SF}$, where
$\tilde{\bmr}_1^\mathrm{SF}$ and $\tilde{\bmr}_2^\mathrm{SF}$ are the
Cartesian coordinates of the oxygen atoms in the reference
geometry. We define a two-angle embedded body-fixed (BF) frame 
$\mathbb{R}(\Phi,\Theta,0)$ in the
$zyz$-Euler angle parameterization (p.~23 of Ref.~\onlinecite{biedenharn:81}),
where $(\Theta,\Phi)$ are the spherical polar angles of $\bmR$,
so that SF coordinates $\bmr_i^\mathrm{SF}$ are related
to BF coordinates $\bmr_i^\mathrm{BF}$ by
\begin{equation}
 \bmr_i^\mathrm{SF} = \mathbb{R}(\Phi,\Theta,0) \bmr_i^\mathrm{BF}
\end{equation}
and the $z$-axis of the BF frame is along the Jacobi scattering
vector. We also define molecule fixed (MF) coordinates through
\begin{equation}
   \bmr_i^{\mathrm{BF}} = 
    \mathbb{R}(\alpha,\beta,0)\bmr_i^{\mathrm{MF}},
\end{equation}
where $(\beta,\alpha)$ are the spherical polar angles
of the vector $\tilde{\bmr}$ in the BF frame.
The MF spherical polar angles of the helium atom are
$(-\beta,0)$, which may also be written as $(\beta,\pi)$.
This is important since we use MF coordinates when computing
the interaction potential.

To describe the vibrations of CO$_2$ we use dimensionless
rectilinear normal coordinates \Qs\ for the symmetric stretch $(\nu_1)$,
\Qx\ and \Qy\ for the degenerate bend $(\nu_2)$,
and $\Qa$ for the asymmetric stretch $(\nu_3)$. Although we
only include the symmetric stretch and bend in the
current study, we define $\Qa$ for completeness and
future reference.
These coordinates are related to the MF Cartesian
displacements $\{(x_i,y_i,z_i), i=1,C,2$\} of the atoms
with respect to the
reference geometry,
\begin{align}
  \Qs \equiv \Half (z_1 - z_2) \sqrt{\mu_1\omega_1/\hbar},\\
  \Qx \equiv (x_1 - x_\mathrm{C}) \sqrt{\mu_2\omega_2/\hbar},\\
  \Qy \equiv (y_1 - y_\mathrm{C}) \sqrt{\mu_2\omega_2/\hbar},\\
  \Qa \equiv \left(\frac{z_1+ z_2}{2} - z_\mathrm{C}\right)
   \sqrt{\mu_3\omega_3/\hbar}
\end{align}
with reduced masses $\mu_1  = 2m_\mathrm{O}$ and
\begin{align}
  \mu_2 = \mu_3 & = \frac{2m_\mathrm{O}m_\mathrm{C}}{2m_\mathrm{O}+m_\mathrm{C}},
\end{align}
with atomic masses 
$m_\mathrm{O} = 15.995$ u
and
$m_\mathrm{C}= 12$ u.
  
We will also make use of the polar coordinates $(\Qb,\gamma)$ of
the bend normal mode, which are related to the Cartesian components
by
\begin{align}
\label{eq:Qxy}
\begin{pmatrix}
  \Qx\\
  \Qy
\end{pmatrix}
= \Qb
\begin{pmatrix}
  \cos\gamma\\
  \sin\gamma
  \end{pmatrix}.
\end{align}
Furthermore, we use complex bend coordinates
\begin{equation}
 Q_\pm\equiv \Qx\pm i \Qy = \Qb e^{\pm i \gamma}.
\end{equation}
\subsection{Hamiltonian and channel basis}
\label{sec:hamiltonian}
The Hamiltonian of the \ce{CO2-He} complex is
\begin{align}
  \hat{H} =& -\frac{\hbar^2}{2\mu_R R}\Ddf{}{R}R
  + \frac{\hat{J}^2+\hat{j}^2 - 2\hat{\bmj}^\mathrm{BF}\cdot\hat{\bmJ}^\mathrm{BF}}{2\mu_R R^2}\nonumber \\
  &\mbox{}+\hat{H}_{\mathrm{CO}_2}^\mathrm{vib}(\Qs,\Qx,\Qy)
  + B\left[\hat{j}^2-\left(\hat{j}_z^\mathrm{MF}\right)^2\right]\nonumber\\
  &\mbox{} +V(\Qs,\Qx,\Qy,\beta,R).
\end{align}
The first term is the radial kinetic energy, with $\mu_R$ the
reduced mass of the complex, the second term is the
centrifugal energy of the complex, where we use BF angular momentum operators
$\hat{\bmj}$ for CO$_2$ and $\hat{\bmJ}$ for the complex.

The third term is the vibrational Hamiltonian of CO$_2$, for which we
use the normal mode expansion of Rodriguez-Garcia {\em et al.}\ \cite{rodriguez:07}.
The harmonic frequencies of CO$_2$ are $\omega_b=680.8$ cm$^{-1}$ for the
bend, $\omega_s=1343.1$ cm$^{-1}$ for the stretch, and $\omega_a=2377.9$
cm$^{-1}$ for the asymmetric stretch. The anharmonic and coupling terms
are essential to properly describe the Fermi resonance and
further details are given in Sec.\ \ref{sec:normal} below.
We do not present cross sections for transitions involving the asymmetric
stretch mode, but we keep it in the discussion of symmetry properties below
for completeness and future reference.

The fourth term is the rotational kinetic energy of CO$_2$, with rotational
constant $B=0.3903$ cm$^{-1}$. The only contribution to the MF component
of the angular momentum is vibrational angular momentum $\hbar l_2$
of the bending mode, so the rotational energies are given by
\begin{equation}
 \epsilon_{jl_2} = B[j(j+1)-l_2^2].
\end{equation}
We do not include centrifugal distortion or rotation-vibration coupling
in CO$_2$.

The final term is the normal mode dependent interaction potential,
which gives rise to the rovibrational transitions. The {\em
ab initio} calculations and the fit are described in the
next section.

The rotational wave function of the complex is
\begin{equation}
  \Ket{\Omega JM} = D^{J,\ast}_{M\Omega}(\Phi,\Theta,0) \sqrt{\frac{2J+1}{4\pi}},
\end{equation}
where $J$ is the total angular momentum quantum number and
$\hbar M$ and $\hbar \Omega$ are the projections
of the total angular momentum on the SF and BF $z$-axes, respectively.
The rovibrational wave function of CO$_2$ is
\begin{equation}
  \label{eq:rovib}
 \Ket{n l_2 j \Omega} = \Ket{n l_2} D^{j,\ast}_{\Omega l_2}(\alpha,\beta,0) \sqrt{\frac{2j+1}{4\pi}},
\end{equation}
where we use $\Ket{nl_2}$ with $n=1,2,\ldots$ as short-hand for
vibrational wave functions $\Ket{\vc_1\vc_2^{l_2}\vc_3(r)}$, sorted by
energy for each value of the vibrational angular momentum $l_2$,
discussed in Sec.\ \ref{sec:normal} below.

Permutation of the oxygen nuclei gives
\begin{equation}
 \hat{P}_{12} \Ket{n  l_2 j\Omega\epsilon} = (-1)^j \Ket{n, {-l_2},j,\Omega,\epsilon}
\end{equation}
so the permutation symmetry adapted CO$_2$ functions are
 \begin{equation}
  \Ket{n\tlv j\Omega\epsilon} =
  \frac{1+\epsilon\hat{P}_{12}}{\sqrt{2+2\delta_{\tlv,0}}}
  \Ket{n\tilde{l}_vj\Omega},
\end{equation}
where $\tilde{l}_2=|l_2|$. Since the $^{16}$O nuclei are bosons we have $\epsilon=1$.

The channel functions are the products of the permutation symmetry adapted
rovibrational CO$_2$ functions and the rotational wave function of the complex.
Under inversion $(\hat{i})$ they transform as
 \begin{equation}
   \hat{i} \Ket{n\tlv j  \Omega \epsilon J M} =
   \epsilon (-1)^{\tlv+j+J}  \Ket{n\tlv j, {-\Omega}\epsilon J M},
\end{equation}
so the parity $(p=\pm 1)$ adapted channel functions are
\begin{equation}
   \Ket{n\tlv j \tOmega \epsilon J M p} =
   \frac{1+p\,\hat{i}}{\sqrt{2+2\delta_{\tOmega,0}}}
   \Ket{n\tlv j  \tOmega \epsilon J M},
\end{equation}
where $\tOmega = |\Omega|$.
\subsection{Five dimensional \ce{CO2-He} potential}
\label{sec:PES}
\subsubsection{{\em Ab initio} method}
The potential was calculated with the explicitly-correlated coupled-cluster
method with single, double, and perturbative triple excitations,
CCSD(T)-F12a using the {\sc Molpro} program package \cite{molpro:15} version 2015.1.
The F12 correction for the triples is included by scaling
the triples energy contribution (the \verb+SCALE_TRIP=1+ option).
We used the augmented correlation-consistent polarized
valence triple-zeta (aug-cc-pVTZ) Dunning basis.
We also apply the Boys and Bernardi \cite{boys:70} counterpoise
correction of the basis set superposition error (BSSE).
\\
\subsubsection{Coordinate grid}
A five dimensional grid was used. For the scattering coordinate the
grid has $n_R=24$ points in the interval $[R_0,R_{n-1}]=[3.25,16]$ $a_0$, with
\begin{equation}
  R_i = R_0 + i\Delta + g_6 i^6,\;\;\;\mbox{$i=0,1,\ldots,n_R-1$},
\end{equation}
where, $\Delta=0.2$ $a_0$ and $g_6$ was chosen such that $R_{n-1}=16$ $a_0$.
For the angle $\beta$ we use $n_{\beta}=16$ Gauss-Legendre quadrature
points in the interval $[0,\pi]$.

For the dimensionless stretch normal coordinate we use $\ns=5$ points,
$\Qs=[-2,-1,0,1,2]$.
The spherical polar angles of the helium atom in the
two-angle embedded MF frame are $(\beta,\pi)$.
In Ref.~\onlinecite{selim:23} a three-angle embedded MF frame
was used for which the azimuthal angle of the helium atom
was $\gamma'=\pi-\gamma$, but here we directly use $\gamma$.
We use a $\nb=5$ point grid $\Qb = [-3, -1.5,0,1.5,3]$
and for the angle $\gamma$ we use $n_\gamma=5$ points
\begin{equation}
  \gamma_i = \pi (2i-1)/(2n_\gamma),\;\;\; \mbox{$i=1,2,\ldots, n_\gamma$}.
\end{equation}
Thus, the primitive grid has
$\ns \nb n_\beta n_\gamma n_R =5 \times 5 \times 16 \times 5 \times 48 = 48\, 000$
points, i.e., 2000 for each $R_i$. However, this grid has redundant points: for the linear geometry,
i.e., for $\Qb=0$, all 5 $\gamma$ points are equivalent, which leaves
$n_{\beta}\ns[1+(\nb-1)n_{\gamma}]=1680$ points.
Furthermore, because of symmetry, we only need to include points with $\beta < \pi/2$,
 which further reduces the number of independent points
to 840 for each value of $R$. Finally, since the potential is invariant
under $(\gamma, Q_b)\rightarrow(\pi-\gamma, -Q_b)$ 520 independent
points for each $R_i$ remain.
\subsubsection{Expansion of the potential}
Since we only require relatively small amplitude vibrations of CO$_2$,
we use a polynomial expansion for the normal mode dependence
of the interaction potential,
\begin{align}
\label{eq:potential}
  V(\Qs,\Qb,\beta,\gamma,&R) =
  \sum_{\ML=0}^{\MLmax}\;
  \sum_{\Pb=\ML}^{\Pbmax(\ML)}\; \sum_{\LL=\ML}^{\LLmax}\;
  \sum_{\Ps=0}^{\Psmax}\\
  &\Qs^\Ps \Qb^\Pb C_{\LL \ML}(\beta,0) \cos(\ML\gamma)\,
  c_{\Ps \Pb \LL \ML}(R),\nonumber
\end{align}
where the functions $C_{\LL \ML}$ are Racah normalized spherical harmonics.
In Appendix \ref{sec:appendixV} we derive the functional
form, the symmetry constraints $(-1)^\ML=(-1)^\Pb=(-1)^\LL$,
the constraint that $\Pb\ge |\mu|$, and we give the summation
limits and further details that result in an almost exact fit to
the {\em ab initio} points.

The largest deviation between the {\em ab initio} points and the
fit is $0.2$ cm$^{-1}$ for the attractive points on the potential.
A {\sc Fortran} implementation of the potential is provided
in the Supplementary Information.
\subsection{Vibrational basis and \ce{CO2} monomer Hamiltonian}
\begin{table}[t]
  \caption{\label{tab:forcefield}Expansion coefficients of \ce{CO2}
    potential in Eqs.~(\ref{eq:Va}) and (\ref{eq:Vc}), taken from
    Ref.~\onlinecite{rodriguez:07}, converted to
    dimensionless normal coordinates, and multiplied by $10^6$.}
  \begin{ruledtabular}
  \begin{tabular}{lr}
                & $10^{-6}$ au\\ \hline
    $\gs$       & $-484.9598$\\
    $\tsss$     & 1219.6662\\
    $\ussss$  & 201.1798\\
    $\ubbbb $  & 295.2919\\
    $\tsbb $   & $-663.3248$\\
    $\ussbb $  & $-200.4589$
  \end{tabular}
  \end{ruledtabular}
\end{table}
\label{sec:normal}
We use a harmonic oscillator basis to describe the symmetric stretch mode $(\nu_1$)
\begin{equation}
  \Ket{\vs} = \frac{1}{\sqrt{\vs!}} (\has^\dagger)^{\vs}\Ket{0},\;\;
  \mbox{with $\vs=0,1,2,3,4$},
\end{equation}
where $\hasd$ is the harmonic oscillator creation operator and $\Ket{0}$
the vibrational ground state for $\nu_1$. The bend could be
described analogously in a Cartesian harmonic oscillator product basis
\begin{equation}
  \Ket{n_xn_y}^{x,y} = \frac{1}{\sqrt{n_x!n_y!}} (\haxd)^{n_x}
  (\hayd)^{n_y}\Ket{0,0},
\end{equation}
but we found it more convenient to use a spherical basis
\begin{equation}
  \Ket{\np\nm}^{+,-} = \frac{1}{\sqrt{\np!\nm!}} (a_+^\dagger)^{n_+}
  (a_-^\dagger)^{n_-}\Ket{0,0}.
\end{equation}
The spherical creation operators are related to the Cartesian ones through
\begin{equation}
  \hapmd =\frac{1}{\sqrt{2}}(\haxd \pm i \hayd).
\end{equation}
The total bend vibrational quantum number is $\Vb =\np+\nm$
and the vibrational angular momentum is given by $\lb =\np-\nm$.
The spherical basis can be converted to a basis using these
quantum numbers
\begin{equation}
  \Ket{\Vb \lb } = \Ket{\frac{\Vb +\lb }{2},\frac{\Vb -\lb }{2}}^{+,-},
\end{equation}
where $\lb =-\Vb ,-\Vb +2,\ldots,\Vb $.
Reflection in the MF $xz$-plane gives
\begin{equation}
   \hat{\sigma}_v(xz)\Ket{\Vb\lb } = \Ket{\Vb ,{-\lb }}.
\end{equation}

We chose a vibrational basis with $\Vb =0,1,\dots,6$, restricting
the vibrational angular momentum to $|\lb |\leq 2$.

We set up the vibrational Hamiltonian of \ce{CO2} in the $\Ket{\np\nm}^{+,-}$
basis, converting it to the $\Ket{\Vb \lb }$ basis at the end. 
It is given by
\begin{equation}
\label{eq:Hvib}
  \hat{H}_\mathrm{vib} = \hbar\omega_s(\hat{N}_s+\Half) + \hbar\omega_b
  (\hat{N}_b+1)+
   \hat{V}_{\ce{CO2}}(\Qs,Q_+,Q_-),
\end{equation}
with number operators
\begin{align}
  \hat{N}_s & = \hasd \has\\
  \hat{N}_\pm & = \hapmd \hapm\\
  \hat{N}_b  & = \hat{N}_++\hat{N}_-.
\end{align}
The potential $\hat{V}_{\ce{CO2}}$ depends on the stretch coordinates
\begin{equation}
  \hQs = \frac{1}{\sqrt{2}}(\hasd + \has)
\end{equation}
and on the spherical bend coordinates
\begin{equation}
  \hQpm = \hQx \pm i \hQy = \hamp + \hapmd.
\end{equation}
The potential contains anharmonic terms and coupling between the modes
\begin{equation}
  \hat{V}_{\ce{CO2}} = \hat{V}_{\mathrm{a}} + \hat{V}_{\mathrm{c}}.
\end{equation}
The anharmonic part is given by
\begin{equation}
  \label{eq:Va}
  \hat{V}_\mathrm{a} = \gs \hQs+\frac{1}{6}\tsss\hQs^3 +
  \frac{1}{24}{\ussss}\hQs^4+\ubbbb  (\hQp \hQm)^2
\end{equation}
and the coupling between the modes is given by
\begin{equation}
  \label{eq:Vc}
  \hat{V}_\mathrm{c} = \frac{1}{6}(\tsbb \hQs+\frac{1}{4}\ussbb \hQs^2)(\hQp\hQm)^2
\end{equation}
To compute matrix elements of the Hamiltonian we first rewrite all the
terms in normal form of products of creation and annihilation operators
using Eq.~(8) of \cite{krasnoshchekov:14}.
This Eq.\ also applies to the
spherical creation and annihilation operators, since the $\hap$ and $\hapd$
operators commute with $\ham$ and $\hamd$, and otherwise satisfy 
commutation relations equivalent to those of $\has$ and $\hasd$.

We took the expansion coefficients in Eqs.~(\ref{eq:Va}) and (\ref{eq:Vc}) from
the work of Rodriguez-Garcia {\em et al.} \cite{rodriguez:07}.
They give the potential in Cartesian normal modes in Eq.~(7) of their paper,
and the coefficients are tabulated in their Table III. We convert  the
potential to spherical coordinates using
\begin{equation}
  Q_+Q_- = (Q_x+iQ_y)(Q_x-iQ_y)=Q_x^2+Q_y^2 = Q_b^2.
\end{equation}
Rodriguez-Garcia {\em et al.}\ use mass-weighted, Cartesian, normal 
coordinates [see Eqs.~(1) and (2) of their paper], with masses in
atomic mass units (u) and distances in \AA. We converted their potential
to our dimensionless coordinates, dropping the dependence on the asymmetric
stretch coordinate.

Considering that the bend terms in the force field depend only on the bend
amplitude $\Qb$, it follows that the quartic bend terms defined in Eq.~(7) of
Ref.~\onlinecite{rodriguez:07} depend on $\Qb^2$, which in Cartesian coordinates reads
$(\Qx^2 + \Qy^2)^2 = \Qx^4 + 2\Qx^2\Qy^2 + \Qy^4$.
This implies that the coefficient $\uxxyy$
must be 1/3 of the coefficient $\uxxxx$. In Table III of
 Ref.~\onlinecite{rodriguez:07} this ratio is
0.12040/0.041718 = 2.8860444, so we adjusted the coefficient $\uxxyy$ to be
precisely 1/3 of $\uxxxx$. The converted and adjusted force field coefficients
are given in Table \ref{tab:forcefield}.

The CO$_2$ anharmonic vibrational wave functions are expanded in
the harmonic basis
\begin{equation}
   \label{eq:nl2}
   \Ket{n l_2} = \sum_{\vc_s=0}^4 \sum_{\vc_b=0}^6
    \Ket{\vc_s \np \nm} c_{\vc_s \np \nm ;n l_2},
 \end{equation}
where $l_b=l_2$ since the vibrational angular momentum is
a good quantum number with our monomer Hamiltonian. The
expansion coefficient $c_{\vc_s \vc_b;\vc_2 l_2}$
are determined variationally. The eigenvalues
and eigenvectors [Table \ref{tab:normal}]
are in very good agreement with the results of Rodriguez-Garcia {\em et al.}

\begin{table}
\caption{\label{tab:normal}
Eigenvalues and eigenvectors of the vibrational Hamiltonian
[Eq.~(\ref{eq:Hvib})] using the parameters of Table \ref{tab:forcefield}.
The coefficients in red are the main components
of the Fermi resonance doublet.
The $(n,l_2)$ labeling of the eigenstates is explained
in Eqs.~(\ref{eq:rovib}) and (\ref{eq:nl2}). The rows correspond to the harmonic basis
functions $\Ket{\vc_s\vc_b^{l_2}\vc_a}$ with non-negligible contributions.}
\begin{ruledtabular}
\begin{tabular}{lrrrrr}
           &  $00^00$ &  $01^10$  &  $10^00 (2)$ & $02^20$ & $10^00 (1)$\\
$(n, l_2)$ &  (1,0)  &  (1,1)  &  (2,0) &  (1, 2)& (3,0)\\
Experiment \cite{rodriguez:07} (cm$^{-1}$)&& 667.4 & 1285.4 & 1335.1 & 1388.2\\
Calculated  & 0  & 670.0  &1288.4 & 1341.3 & 1391.3\\
\hline
$\Ket{00^00}$ &  1.000 & 0.000 & $-0.001$ & 0.000 & $-0.011$\\
$\Ket{01^10}$ &  0.000 & 1.000 &   0.000  & 0.000 &   0.000 \\
$\Ket{10^00}$ &$-0.000$& 0.000 & \Red{0.721}  & 0.000 & \Red{$-0.690$}\\
$\Ket{02^20}$ &  0.000 & 0.000 &   0.000  & 0.997 &   0.000 \\
$\Ket{02^00}$ &  0.007 & 0.000 & \Red{0.680}  & 0.000 & \Red{0.729}
\end{tabular}
\end{ruledtabular}
\end{table}
\subsection{Channel basis matrix elements}
\label{sec:matrixelements}
To compute matrix elements of the potential, we use
\begin{align}
  \Bra{l_v' &j'\Omega' }C_{\lambda\mu}(\beta,\gamma) \Ket{l_v  j  \Omega } =\\
  &(-1)^{l_v-\Omega} [j',j]^{\Half}
  \Threejm{j'}{-l_v}{\lambda}{\mu}{j}{l_v}
  \Threejm{j'}{-\Omega'}{\lambda}{0}{j}{\Omega}\nonumber,
\end{align}
together with Eq.~(\ref{eq:cos}) in the Appendix.
These matrix elements are nonzero only for $l_v'=\mu+l_v$.

To compute matrix elements over the vibrational coordinate
dependence we employ the bend coordinate dependence
expressed in spherical components
\begin{equation}
  \Qb^\Pb e^{i\mu\gamma} = \hQp^\Pp\hQm^\Pm
\end{equation}
and use the potential expanded in spherical normal coordinates
\begin{align}
  &\Bra{n'l_2'}\Qs^{\Ps} \Qb^\Pb e^{i\mu\gamma}\Ket{nl_2} =
  \sum_{\vc_s'\np'\nm'} c^\ast_{\vc_s'\np'\nm';n'l_2'}\nonumber\\
  &\times \sum_{\vc_s\np\nm}
    \Bra{\vc_s'\np'\nm'}\hQp^\Pp\hQm^\Pm\Ket{\vc_s \np\nm}
  c_{\vc_s\np\nm;nl_2}
\end{align}

The matrix elements over the BF rotational energy terms
are calculated in the usual way, see, e.g., Ref.~\onlinecite{dhont:04}.
\subsection{Energy grid and selection of multiple initial rotational states}
\label{sec:Egrid}
Converging collisional rate coefficients for temperatures between 100-800 K requires
about  $n_\mathrm{coll}=30$ collision energies in the
range of $E_\mathrm{min}^\mathrm{coll}=20$ cm$^{-1}$ to
$E_\mathrm{max}^\mathrm{coll}=5220$
cm$^{-1}$ on a logarithmic grid,
\begin{equation}
\label{eq:Ecoll}
  E_k^\mathrm{coll} = f^{k-1} E_\mathrm{min}^\mathrm{coll},\;\;\;
  \mbox{$k=1,2,\ldots, n_\mathrm{coll}$, and $f=1.21$}.
\end{equation}
For a given initial state with rovibrational energy $\epsilon_i$ this
requires solving the CC equations at $n_\mathrm{coll}$ different total
energies $E^\mathrm{tot}_{i,k} = \epsilon_i+E_k^\mathrm{coll}$. Thus,
for $n_\mathrm{init}$ initial states, a total of $n_\mathrm{init}\times
n_\mathrm{coll}$ CC calculations would be needed.
Each calculation
yields a full scattering matrix, so cross sections could be computed
not only for initial state $(i)$ with collision energy $E_k^\mathrm{coll}$,
but for all initial states
that are open at the given total energy.
However, the cross sections
for the other initial states $(j)$ would correspond to collision
energies $E^\mathrm{coll}= E_{i,k}^\mathrm{tot}-\epsilon_j$
that are not necessarily on the predefined collision energy grid.

We reduce the number of CC calculations by accepting
slightly different collision energies for different
initial states. From a single CC calculation we use
the initial states with energies $\epsilon_l$ and
$\epsilon_u$ $(>\epsilon_l)$ and all initial states with energies
in between.
Thus, the largest
collision energy difference corresponding to these initial
states is $\Delta E = \epsilon_u-\epsilon_l$.
In Appendix \ref{app:Egrid} we present an algorithm
to recursively generate a total energy grid such that
this difference is at most a fraction $s=0.1$ of
the predefined collision energies for each initial state,
\begin{equation}
\label{eq:DeltaE}
  \Delta E \leq s E_k^{\mathrm{coll}}
\end{equation}
and the collision energy deviates at most a factor
$\sqrt{1+s}\approx 1.05$ from the target.
Our algorithm is most useful at high collision energies,
where there are many open channels and where the
cross sections vary smoothly with collision energy.
\subsection{Convergence parameters and further details}
\label{sec:conv}

The convergence parameters are the maximum value of the \ce{CO2}
rotational quantum number, $j \leq j_\mathrm{max}$, the maximum 
value of the total angular momentum of the partial waves,
$J \leq J_\mathrm{max}$, the maximum value of the BF projection
quantum number, $|\Omega| \leq \Omega_\mathrm{max}$, and the range
$[R_a,R_b]$ and step size $\Delta_R$ of the renormalized
Numerov propagation grid. We take $R_a=3.25$ $a_0$, $R_b=25$ $a_0$
and we take $\Delta_R=\lambda_D/n_\mathrm{grid}$, where
$\lambda_D$ is the De Broglie wavelength of the open channel
with the largest kinetic energy and $n_\mathrm{grid}=6$.
The values of $j_\mathrm{max}, J_\mathrm{max}$, and $\Omega_\mathrm{max}$
required for convergence depend on the initial rotational state, $j_i$,
and on the collision energy $E_\mathrm{coll}$. We optimized convergence parameters
in test calculations for $j_i=1,5,10,20,$ and 35,
and for collision energies $E_\mathrm{coll}=20, 200, 2000,$ and 5220
cm$^{-1}$. In the test calculations we 
only included every 5th partial wave, $J=0,5,10,\ldots$ to
reduce the cost, but in the production runs we included all partial waves.
Convergence parameters for other values of $j_i$ and $E_\mathrm{coll}$
were obtained by two-dimensional spline interpolation of the
parameters determined in the test calculations.

For the highest collision energy, $E_\mathrm{coll}=5220$ cm$^{-1}$, and
highest initial rotational state, $j_i=35$, we
need $j_\mathrm{max}=120$ and  $J_\mathrm{max}=118$.
In the BF channel functions one must
have $|\Omega|\leq \min(j,J)$. We found that it is sufficient
to take $\Omega_\mathrm{max} = j_i+2$, even if there are
basis functions with $j\gg j_i$.

The transitions involving the Fermi resonance states
are the most expensive and require
up to 9674 coupled channels. The full data set required
a total of 735\,514 CC calculations.
The number is large since calculations for different
total energies, or different partial waves ($J$),
or different parities, are all independent. Without
the total energy grid optimization the number would
be about an order of magnitude larger. Jobs
were run in parallel on a local linux
cluster and 240 linux desktops available to us outside
office hours. The quantum scattering code was written
in Scilab and the parallel jobs were managed using a
PostgreSQL database.
\section{Results and discussion}
\label{sec:results}

\subsection{CC results and propensity rules}
\label{sec:results-prop}
\begin{figure*}[!tp]
    \centering
    \includegraphics[width=16cm]{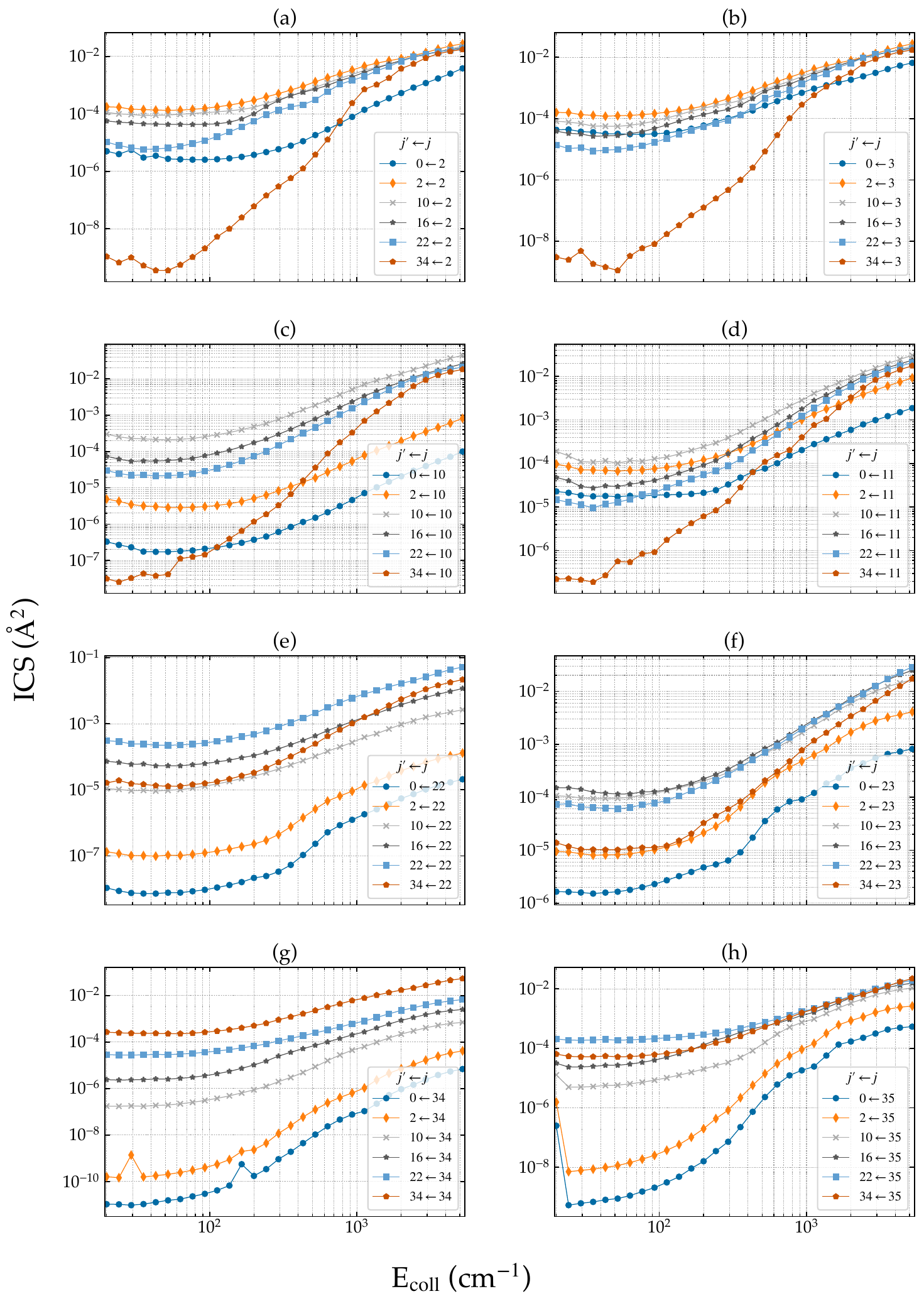}
    \caption{State-to-state integral cross sections as functions of collision
    energy, for the rovibrational transitions $00^00,j\prime\leftarrow01^10,j$.
    Panels show results for initial rovibrational states
    $j=2,\,3,\,10,\,11,\,22,\,23,\,34$ and 35. In each panel, cross
    sections/rate coefficients are shown for transitions to five final
    rotational states: $j=0,\,2,\,10,\,16,\,22$ and $34$.}
    \label{fig:results-cc-1}
\end{figure*}

\begin{figure*}[!tp]
    \centering
    \includegraphics[width=\hsize]{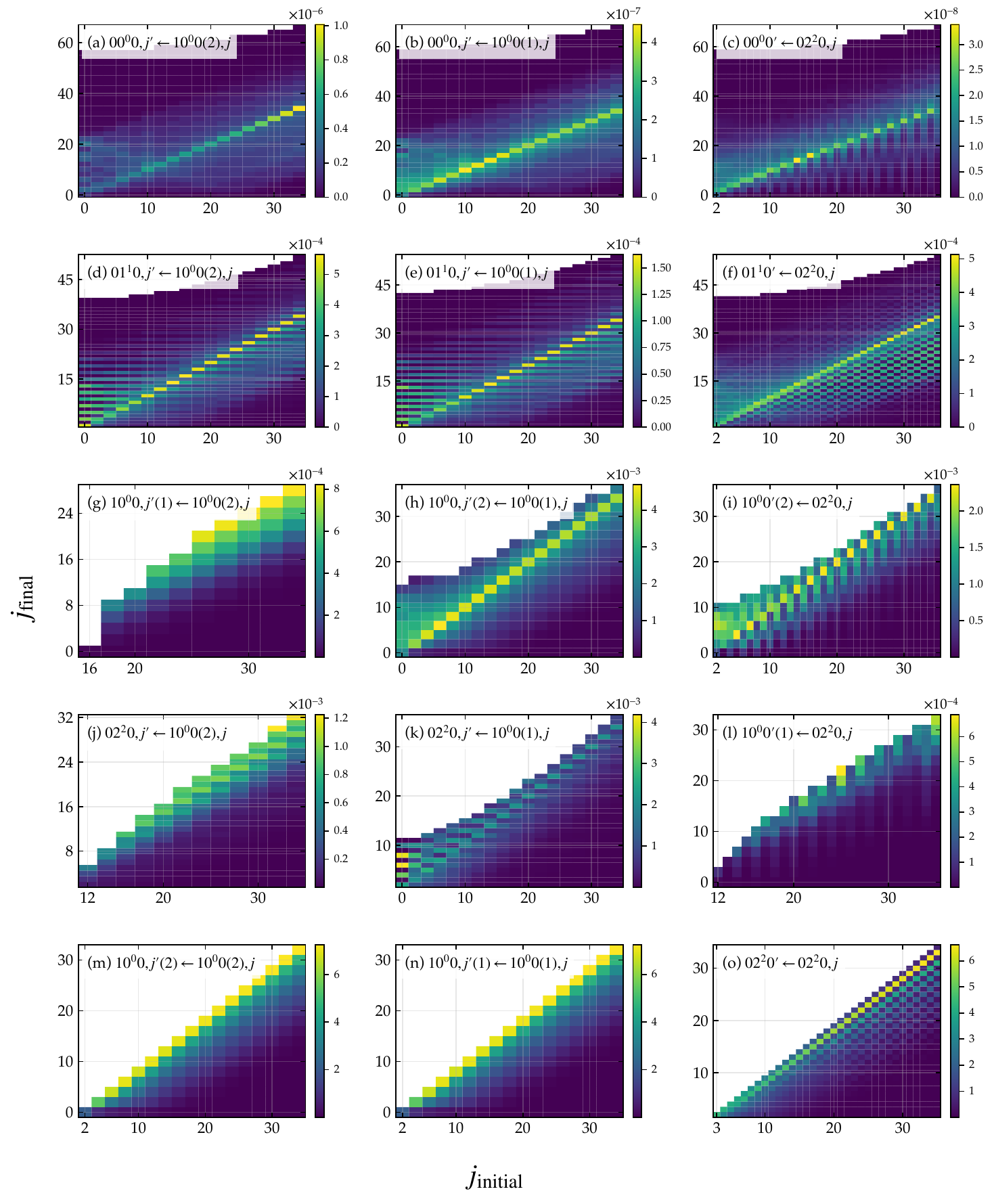}
    \caption{Two-dimensional color maps of the integral cross sections
    [$\mathrm{ICS}_{j'\leftarrow j}$$(\AA^{2})]$ as functions of the initial
    ($j_{\mathrm{initial}}$=$j$) and final ($j_{\mathrm{final}}$=$j'$)
    rotational states for different rovibrational transitions at 200 cm$^{-1}$.
    The first column 
    shows transitions from the $10^00(2)$
    initial states, the second column 
    from $10^00(1)$, and the
    third column 
    from $02^20$. In each column, the last row,
    panels (m-o), correspond to purely
    rotational transitions, where the initial and final vibrational states are
    identical.}
    \label{fig:results-cc-2}
\end{figure*}

\begin{figure*}[!tp]
    \centering
    \includegraphics[width=14cm]{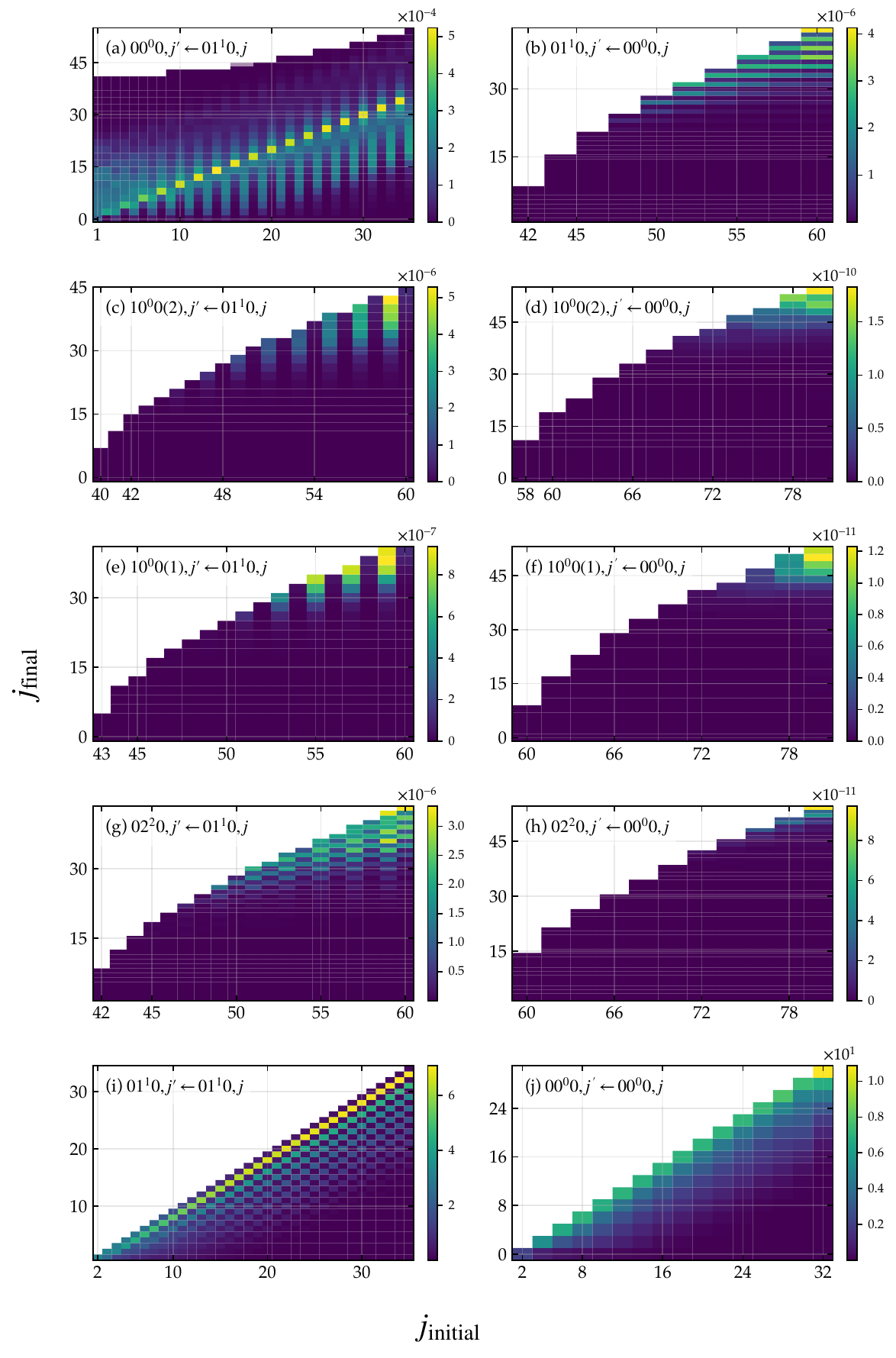}
    \caption{As Fig.\ \ref{fig:results-cc-2}, with the first
    column 
    showing transitions from the bend ($01^10$) initial
    state and the second column 
    from the ground ($00^00$)
    initial state.
    The last row, panels (i) and (j) show pure rotational
    transitions in the first excited bend and the vibrational ground state, respectively.} \label{fig:results-cc-3} \end{figure*}

\begin{figure*}[!tp]
    \centering
    \includegraphics[width=\hsize]{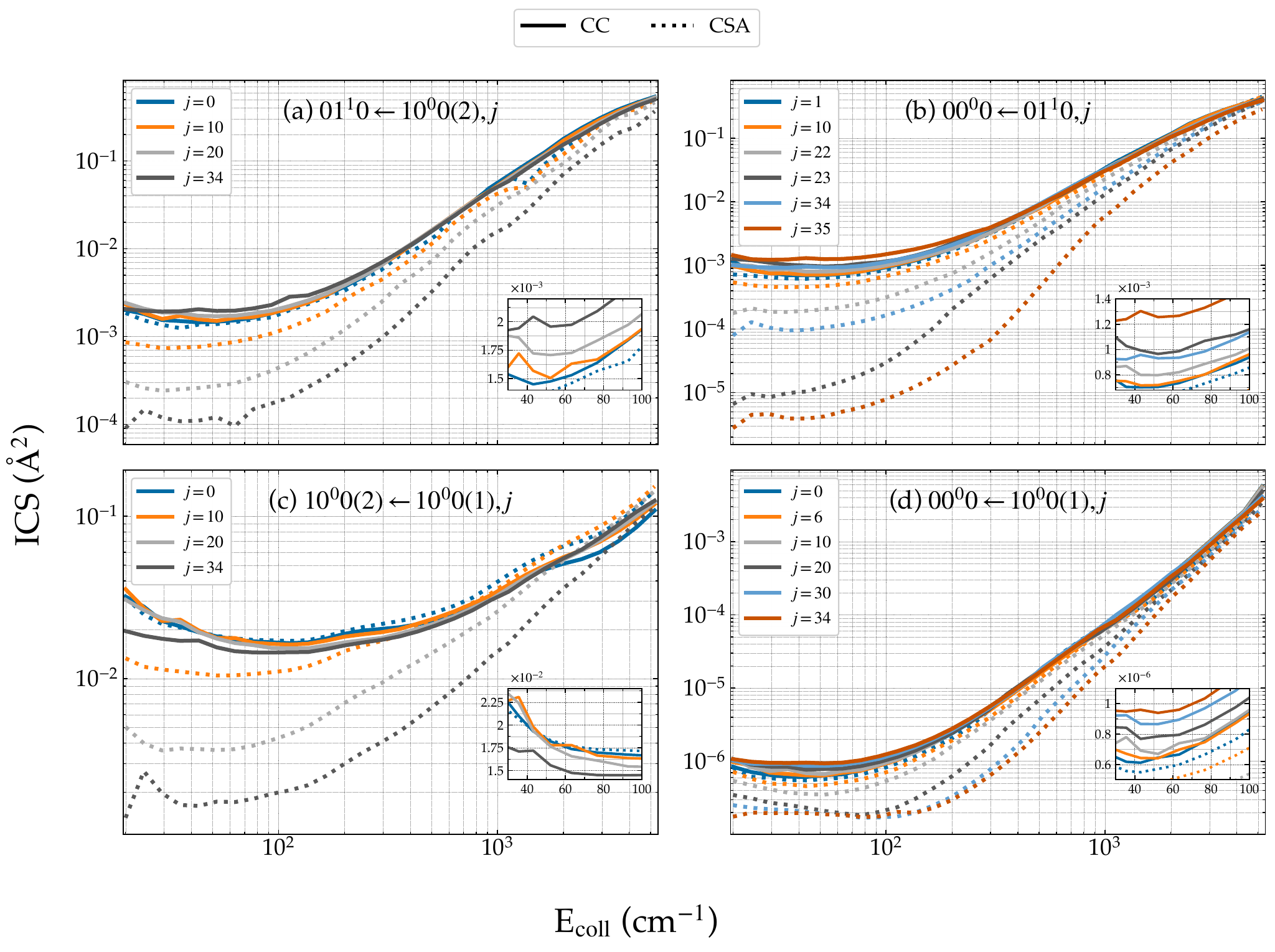}
    \caption{Comparison of CC (solid lines) and CSA (dotted lines) quenching
    integral cross sections summed over all final states $j'$, for different
    rovibrational transitions: 
    (a) $01^10\leftarrow10^00(2)$,
    (b) $00^00\leftarrow01^10$,
    (c) $10^00(2)\leftarrow10^00(1)$, and 
    (d) $00^00\leftarrow 10^00(1)$.
    The CSA results depend strongly on the initial $j$. The insets show
    for collision energies below 100 cm$^{-1}$ that the CC cross sections
    for the $j=34/35$ are higher than for the lowest $j$ by up to a
    factor of two, except for the upper Fermi level, panel (c),
    where they are smaller.}
    \label{fig:results-cc-csa-1}
\end{figure*}

\begin{figure}
    \centering
    \includegraphics[width=0.9\hsize]{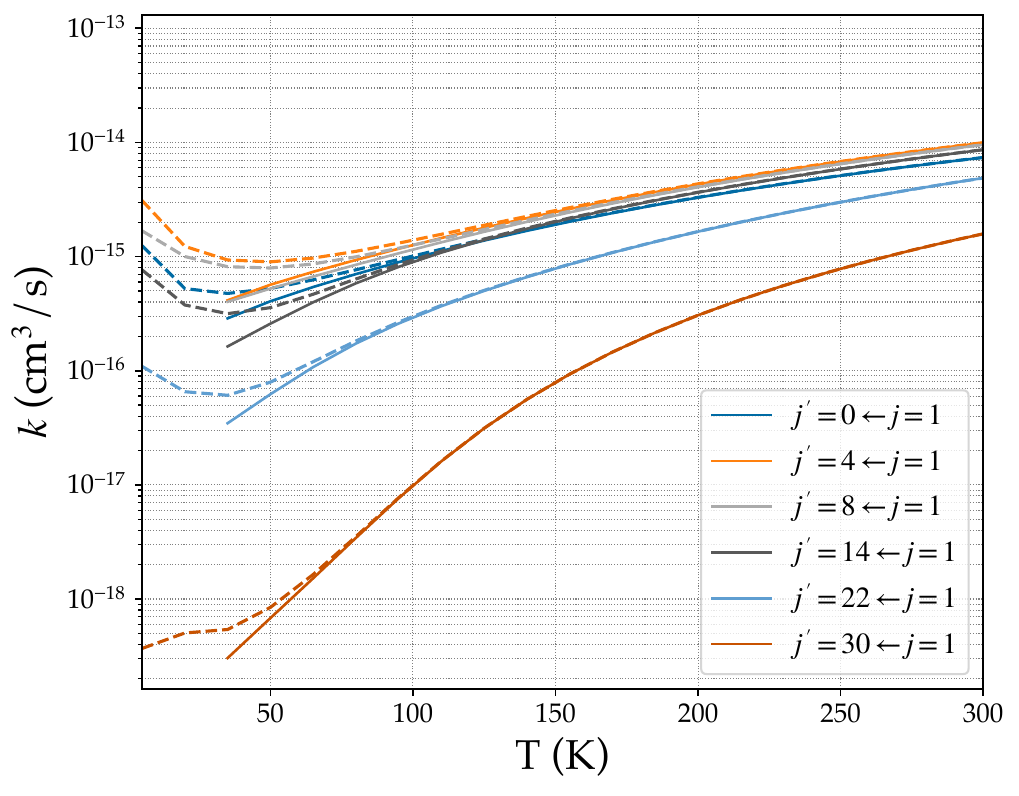}
    \caption{Rate coefficients for the
    $00^00,j'\leftarrow01^10,j=1$  transitions
    as a function of temperature. Dashed lines represent calculations including
    low-energy resonances (1-2000 cm$^{-1}$), while solid lines show results
    excluding resonances (20-5220 cm$^{-1}$).}
    \label{fig:conv-rates}
\end{figure}

\begin{figure}
    \centering
    \includegraphics[width=0.9\hsize]{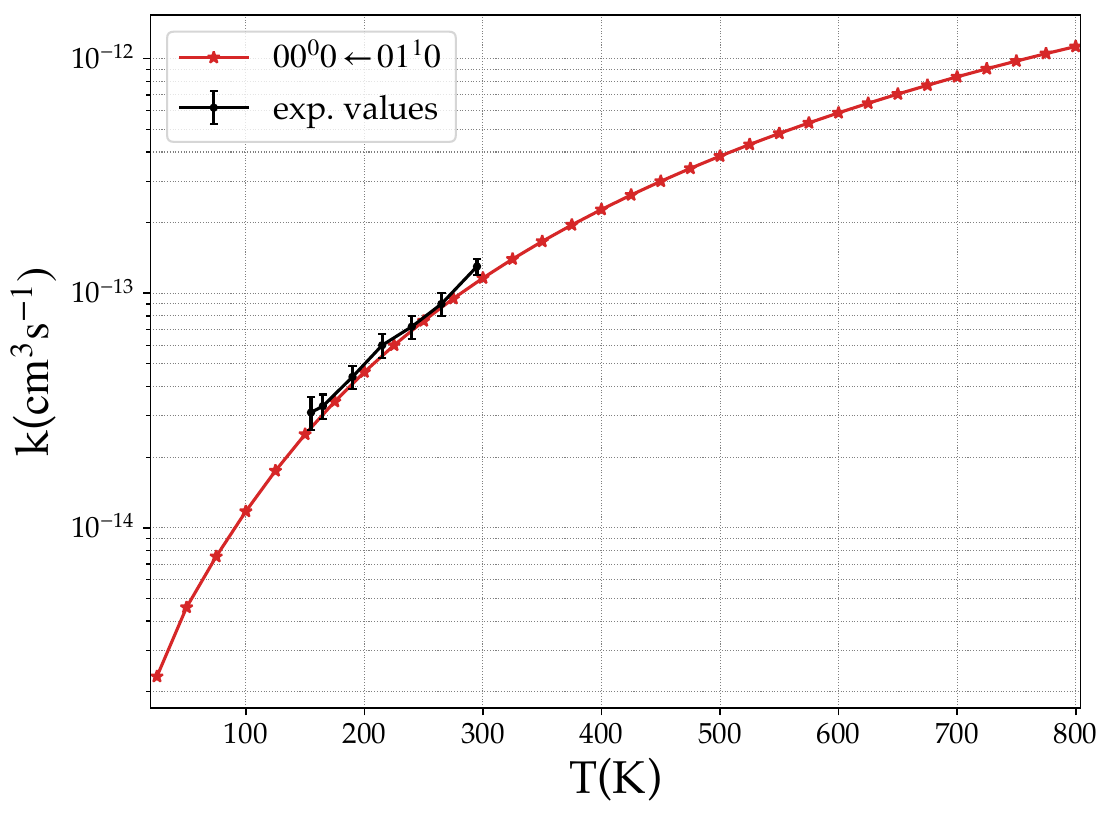}
    \caption{
    Vibrational quenching rate coefficients of the bend fundamental computed using
    the cross sections computed in this study,
    compared with experimental values.
    \label{fig:results-cc-rates-thermal}}
\end{figure}

Figure \ref{fig:results-cc-1} presents the calculated integral cross
sections (ICSs) as functions of collision energy for the rovibrational
transitions $\vc_1\vc_2^{l_2}\vc_3(r),j: 00^00,j' \leftarrow 01^10,j$. In this
case, only even $j'$ states are present in the ground vibrational state,
whereas both even and odd $j$ states occur in the fundamental bending state.
Panels (a, c, e, g) show a clear $\Delta j=0$ propensity, with the largest ICSs
obtained for transitions with $\Delta j=0$. In panels (a,
b, c, d), transitions to the highest final rotational state, $j'=34$, become
larger than those to $j'=0$ and $j'=2$ at higher collision energies, indicating
that the energy gap rule plays an increasingly prominent role in this regime. In general, the ICSs
decrease as the transition moves further away from $\Delta j=0$ and as the
energy gap increases.

The decrease of the cross sections with increasing $\Delta j$ is related
to the anisotropy of the potential, as quantified by the $\lambda$
dependence of the coefficients in the expansion in Eq. (20). The terms
with $\lambda = 0$ form the isotropic potential, the anisotropic terms
with $\lambda > 0$ couple basis functions with different $j$. The
magnitude of the expansion terms decreases with increasing $\lambda$,
which explains the decrease of the cross sections with increasing
$\Delta j$.

By comparing the parity dependence for even and odd
initial states, we find that, for the same final states, the ICSs are higher
for odd initial states in most cases. The main exception is the $\Delta j=0$
transitions, for which the ICSs are higher for even initial states than for
odd initial states.

The sharp features observed in Fig. \ref{fig:results-cc-1} at collision 
energies below 200 cm$^{-1}$ for the smallest cross sections (from 
10$^{-6}$ to below 10$^{-10}$ \AA$^2$) are likely numerical artefacts. 
We obtained and analyzed similar artefacts in previous studies\cite{zhang:21, avoird:26}, where we found that when calculating the cross sections at nearby energies the sharp features disappeared. In
any case, their effect on our calculated rate coefficients is negligible.

To examine parity-dependent propensity rules in more detail,
Figs.~\ref{fig:results-cc-2} and \ref{fig:results-cc-3} show 2D state-to-state CC
integral cross section color maps as functions of initial and final rotational
states for all 25 combinations of transitions. The ICSs generally
increase for higher initial and final rotational states. Several characteristic
patterns are observed: For $l_2'=0 \leftarrow l_2=0$ transitions,
Fig.~\ref{fig:results-cc-2} panels 
(a, b, h) show diagonal structures with
approximately symmetric decreases away from $\Delta j=0$,
except for transitions where the initial vibrational level lies below the final
one (Fig.~\ref{fig:results-cc-2} panels 
g, h, l
and Fig.~\ref{fig:results-cc-3}
panels b-h).

For transitions $l_2'=1,2 \leftarrow l_2=0$, where the initial state
contains only even rotational levels, horizontal patterns appear in the ICSs
color maps (Fig.~\ref{fig:results-cc-2} panels 
d, e, k). Conversely, for
transitions $l_2'=0 \leftarrow l_2=1,2$, where the final state contains only
even rotational levels, vertical patterns are observed 
(Fig.~\ref{fig:results-cc-2} panels 
c, i and Fig.~\ref{fig:results-cc-3} panel a).

For transitions $l_2'=2 \leftarrow l_2=1$ and $l_2'=1 \leftarrow l_2=2$,
some parity combinations show similar ICS patterns along the $\Delta j=0$
diagonal: $j_{\mathrm{odd}}' \leftarrow j_{\mathrm{even}}$ follows
$j_{\mathrm{even}}' \leftarrow j_{\mathrm{odd}}$, while $j_{\mathrm{odd}}'
\leftarrow j_{\mathrm{odd}}$ follows $j_{\mathrm{even}}' \leftarrow
j_{\mathrm{even}}$ (Fig.~\ref{fig:results-cc-2} panel
f and
Fig.~\ref{fig:results-cc-3} panel
g). Rotational transitions are shown in the last
rows of Fig.~\ref{fig:results-cc-2} (panels
m-o) and
Fig.~\ref{fig:results-cc-3} (panels
i, j), where the largest ICSs occur for $\Delta
j=2$, and decrease toward lower $j$ and $j'$ values. 

Overall, the state-to-state ICSs strongly depend on the parity of the initial
and final rotational states and exhibit clear propensity rules dominated by
$\Delta j=0$.
\subsection{Comparison of integral cross sections: CSA vs CC}
\label{sec:cc/csa}

Figure~\ref{fig:results-cc-csa-1} compares the exact CC and CSA ICSs for
vibrational quenching, summed over all final rotational states $j'$, across a
broad range of initial $j$ states, for several vibrational transitions. The
exact CC ICSs, represented by solid lines, increase by a factor of about
1.5-2 with increasing initial rotational state $j$.
An exception is observed in panel (c), which corresponds
to transitions between Fermi-coupled vibrational levels and where
the highest $j$ gives a smaller cross section at lower collision energies.

By contrast, the CSA cross sections, shown by dotted lines, decrease with
increasing $j$. As a result, the discrepancy between the CC and
CSA results becomes significantly more pronounced at higher $j$ values. In
particular, the CC/CSA differences range from less than a factor of 1.5 at low
initial $j$ to about a factor of 10 at high initial $j$, i.e. they remain
within roughly one order of magnitude over the range $j=0-34$. This trend
must be related to the neglect of off-diagonal Coriolis coupling terms in
the CSA approximation. In the full CC treatment, these terms couple different
$\Omega$ projections of the CO$_2$ rotational angular momentum onto the
intermolecular axis through angular-momentum coupling terms involving $J$ and
$j$. The differences between the CC and CSA ICSs decrease at higher
collision energies, from about a factor of 7.5 to about a factor of 1.5, and
therefore remain within one order of magnitude.

Some of the trends in the difference between CC and CSA were already found
in the study of the bend vibration \cite{selim:23}, but for the lower
initial $j$ investigated there, the differences were less than 50\%.
\subsection{Effect of resonances on rate coefficients}
\label{sec:resonances}
Figure \ref{fig:conv-rates} shows state-to-state temperature dependent rate coefficients
for the $j=1$ rotational state of the bend fundamental to
several rotational states of the vibrational ground state.
The solid lines are computed by Boltzmann averaging the cross sections
of the present work using Eq.~(25) of Ref.~\onlinecite{selim:23}.
The cross sections for the 20-5220 cm$^{-1}$
collision energy range were spline interpolated and the integral
was done with a 1000 point trapezoidal rule. The dashed lines
are obtained by extending the energy down to 1 cm$^{-1}$
using the cross sections from Ref.~\onlinecite{selim:23}.
For collision energies below 20 cm$^{-1}$ there are many strong
resonances, which clearly contribute to  the collisional rate coefficients below
$T=50$ K, but have a negligible effect above $T=100$ K.

We compute rovibrational quenching rate coefficients of the bend fundamental
by summing over final rotational states $j'$ and
Boltzmann averaging over initial $j$, as in Eq.~(4) of Ref.~\onlinecite{song:15b}.
In Fig.~\ref{fig:results-cc-rates-thermal} we compare these
rate coefficients with the experimental values of Siddles {\em et
al.} \cite{siddles:94}. The calculated quenching rate coefficients show good
agreement with experimental data over the 155-295 K range, staying 
mostly within the experimental error bars.

\section{Conclusions}
\label{sec:conclusion}
Rotational and rovibrational state-to-state integral cross sections
for \ce{CO2-He} were computed 
at the CC level for the vibrational ground state, the fundamental bend,
$01^10$, its overtone, $02^20$, and the upper and lower Fermi levels, $10^00(1)$
and $10^00(2)$. The calculations cover all transitions among these
states, giving 20 vibrational transition sets and 5
rotational transition sets. For transitions to lower lying vibrational
states we include initial rotational levels up to $j=35$, and
for transitions to higher lying vibrational states initial rotational
states up to $j=68$.

For the bend fundamental we also compute collisional rate coefficients. They agree
well with experimental rate coefficients \cite{siddles:94} and with
previous CC results \cite{selim:23}.

We also compute vibrational quenching rate coefficients at the CC and CSA level
of theory. For low initial rotational levels and at high
collision energies, the CSA agrees quite well with the CC results.
However, at low collision energies and high initial $j$, 
the CSA may be lower than CC by more than an order of magnitude.

The cross sections show clear propensity rules for rovibrational de-excitation,
including parity-dependent patterns and systematic trends with $\Delta j$.
In a separate study\cite{boskovic:26b} we use the propensity rules when
extrapolating the calculated collisional rate coefficients to higher rotational states.

For rotational transitions, rate coefficients for collisions with helium are sometimes
scaled to estimate rate coefficients for collisions with H$_2$ \cite{tak:20}, which
is the dominant collider in many astrophysical environments.  Experimental
vibrationally inelastic cross sections for \ce{CO2-He} and \ce{CO2-H2} reported
by Stephenson {\em et al.}\cite{stephenson:71}, however, showed
that they can differ by more than an order of magnitude, with distinct
temperature trends. In particular, the \ce{CO2-H2}/\ce{CO2-He} ratio decreases
from about 30 at 300 K to about 3 at 800 K, meaning there is no constant
scaling factor between \ce{CO2-He} and \ce{CO2-H2}.  Therefore, helium cannot
serve as a proxy for H$_2$ in rovibrational transitions, which was also
concluded for collisions involving \ce{H2O} \cite{wiesenfeld:22}.

Nevertheless, the simpler \ce{CO2-He} system 
let us investigate converged rovibrational transitions for high
collision energies and highly excited rotational states of five
vibrational states of \ce{CO2}, including Fermi resonance states
and assess the accuracy of the CSA.
The methodological advancements presented in the work, such as
the algorithm to generate an efficient total
energy grid, will also be useful in \ce{CO2-H2} calculations.

Temperature dependent collisional rate coefficients computed from the \ce{CO2-He}
cross sections presented here will be published elsewhere \cite{boskovic:26b}.
\section{Supplementary Material}
\label{sec:supp}
The supplementary material contains a Fortran implementation of the five
dimensional \ce{CO2-He} potential and the energy dependent cross
sections computed at the CC and CSA levels of theory in
computer readable form for all transitions discussed in this paper.
\section{ACKNOWLEDGMENTS}
We thank Ewine van Dishoeck and Inga Kamp for
valuable discussions and feedback. We thank the C\&CZ IT department
of the Faculty of Science of the Radboud University for support and
making 240 linux PCs available outside office hours. 
The research was funded by The Netherlands
Organisation for Scientific Research, NWO, through the Dutch Astrochemistry
Network III (DAN-III).
\appendix 
\section{Fit of the potential}
\label{sec:appendixV}
To derive the functional form of the
potential expansion, we first consider a helium atom with
spherical polar angles $(\beta,\chi)$ in the MF frame, before
we set $\chi=\pi$. Thus, we can use 
Racah normalized spherical harmonics $C_{\LL \ML}(\beta,\chi)$
in the expansion.
To preserve cylinder symmetry
we need homogeneous potentials $\Qx^{\Px}\Qy^{\Py}$ with
nonnegative powers $\Px$ and $\Py$ with $\Px+\Py\leq \Pb$.
Switching to spherical coordinates $\Qpm$, the $(\beta,\chi)$
dependence can be expanded in functions
\begin{equation}
  \Qp^\Pp\Qm^\Pm C_{\lambda\mu}(\beta,\chi) = \Qb^\Pb
  C_{\LL\ML}(\beta,0) e^{i[(\Pp-\Pm)\gamma + \ML \chi]},
\end{equation}
where $\Pb=\Pp+\Pm$ and $\gamma$ is the polar angle of the
bend coordinate [Eq.~(\ref{eq:Qxy})].
Since we need functions that are invariant under rotation
around
the MF $z$-axis we must have
\begin{equation}
 \Pp - \Pm + \ML = 0.
\end{equation}
If $\Pb\equiv \Pp+\Pm <|\ML|$ then $(\Pm-\Pp)$
cannot be equal to $\ML$, so we must have $\Pb\geq |\ML|$.
Also, if $\ML=\Pm-\Pp$ is odd (even), then $\Pm+\Pp=\Pb$ must
also be odd (even), so we have the constraint
\begin{equation}
  \Pb=|\ML|,|\ML|+2, |\ML|+4, \ldots
\end{equation}
Setting $\chi=\pi$ gives
\begin{equation}
\Qp^\Pp\Qm^\Pm C_{\lambda\mu}(\beta,\pi) = (-1)^\mu \Qb^\Pb C_{\lambda,{-\mu}}(\beta,\gamma).
\end{equation}
Since the potential must be real, we restrict the summation
to nonnegative $\mu$ using
\begin{equation}
\label{eq:cos}
  C_{\lambda\mu}(\beta,0)\cos(\mu\gamma) = \frac{C_{\lambda\mu}(\beta,\gamma)
  +(-1)^\mu C_{\lambda,{-\mu}}(\beta,\gamma)}{2}.
\end{equation}
Reflection in the MF $xy$-plane gives 
\begin{equation}
 \hat{\sigma}(xy)C_{\LL\ML}(\beta,\gamma) 
   = (-1)^{\LL+\ML} C_{\LL\ML}(\beta,\gamma)
\end{equation}
and $\Qa$ changes sign. Since the potential must be
invariant under this symmetry operation,
$\Pa+\LL+\ML$ must be even.
With these constraints, dropping the asymmetric stretch,
we obtain the expansion of the potential in Eq.~(\ref{eq:potential}).

We have five $\gamma$ points, so the 
$\gamma$-dependence can be fit exactly with $\MLmax=4$.
For $\ML=0$, power $\Pb $ must be even, which leaves three degrees of
freedom on our 5 point $\Qb$ grid, which can be fit exactly
with $\Pb =0, 2, 4$, i.e., $\Pb ^\mathrm{max}(0)=4$.
For $\ML>0$ we also have $\Pb >0$, so $\Qb^{\Pb }$ is zero for
$\Qb=0$, and there are only two degrees of freedom left on
our 5 point $\Qb$ grid, so we take $\Pb ^\mathrm{max}(\ML)=\ML+2$ for
$\ML>0$. Thus, for the ($\Qb,\gamma$)-dependence this gives 3+2+2+2+2=11
parameters, while the $5\times 5=25$ point $(\Qb,\gamma)$ coordinate grid has
13 degrees of freedom when taking into account the invariance
of the potential under the transformation $(\Qb,\gamma)\rightarrow (-\Qb,\pi-\gamma)$
[see Eq.~(\ref{eq:Qxy}), with the helium atom in the $xz$-plane].
At first, it may seem surprising that the functional form cannot have
the same number of degrees of freedom as the coordinate grid, but something
similar happens when expanding a function given on a direct product
grid of spherical polar coordinates in spherical harmonics\cite{beentjes:15}.

The $\beta$-grid has 8 points, so for $\ML=0$ we take $\LLmax=15$,
which gives 8 degrees of freedom in the expansion, since $\LL+\ML$ must be even.
We take $\LLmax$ the same for all $\ML$, which gives 8 values for
$\LL$ when $\ML \leq 1$, 7 for $\ML=2,3$, and 6 for $\ML=4$.
Taking into account the possible values for $\Pb $ and $\LL$ for each value
of $\ML$ we count 80 parameters of the $(\Qb,\beta,\gamma)$-dependence.

For $\Qs$ we have 5 points so we take $\Psmax=4$,
so for each $R$ we have $5\times 80=400$ parameters, while
there are 520 coordinate grid points for each $R$.

The expansion coefficients $c_{\Ps \Pb \LL \ML}(R_i)$ 
were found by weighted linear least squares fits for each of the 24 $R_i$ values
separately. For the weights, we used the square-roots of the Gauss-Legendre
quadrature weights. In the first step we fit the $\gamma$-dependence by the
expansion in $\cos \ML\gamma$. Next, for each value of $\ML$ and each value of
$\Qs$ we fit the $(\Qb,\beta)$ dependence. In the last step we fit the $\Qs$
dependence.

For small values of $R$ near the linear geometry of the complex,
the potential $V$ becomes very large. To avoid fit problems, we
damp the potential when it is larger than $V_0=0.1$ \Eh, such that
it remains below 0.2 \Eh, using
\begin{equation}
  V_\mathrm{damped} = V_0 [1+\tanh( V/V_0-1)].
\end{equation}

Next, the expansion coefficients $c_{\Ps \Pb \LL \ML}(R)$ are
fitted by the reproducing kernel Hilbert space (RKHS) method
\cite{ho:96,ho:00}. Two parameters are used in the RKHS method:
the smoothness parameter $n$ and the parameter $m$ that
determines the decay of the potential as $1/R^{m+1}$ beyond the
largest point in the radial grid. We take $n=2$, and
since the dispersion interaction between
\ce{CO2} and He decays as $1/R^6$, we take $m=5$.
This produces a set of RKHS coefficients from which the
expansion coefficients $c_{\Ps \Pb \LL \ML}(R)$ can be
calculated for any value of $R$.

For $R\leq 3.65$ $a_0$ the {\sc Molpro} calculations did not
converge for a few points, in particular near the linear geometry.
Since we fit the $(\Qb,\beta,\gamma)$ 
points together, with somewhat fewer fit parameters than grid
points (80 vs 104), we did not need to adapt the fit procedure
to accommodate for the missing {\em ab initio} points,
with one exception. For the shortest distance, $R=3.25$ with \ce{CO2}
linear, $Q_b=0$, and $Q_s=1$, a single point was missing for
the smallest value of $\theta$. Since only terms in the
potential with $\mu=0$ contribute for linear \ce{CO2} the
expansion is exact. This means that the linear least squares fit
will evaluate to zero for this missing point, introducing
extreme oscillations. We solved this by setting the potential
equal to the maximum damped value, $0.2$ E$_h$, since the potential
is very repulsive here. We only discovered this
issue after we had finished the scattering
calculations. A few test calculations showed that this
only had a noticeable effect on the largest $\Delta j$ transitions,
with very small cross sections, and we decided not to recompute
the entire dataset.
\section{Algorithm for generating a total energy grid}
\label{app:Egrid}
The algorithm generates a set of total energies $E_k^\mathrm{tot}$
for which to solve the CC equations. For each total energy it specifies
for which initial states
the cross sections need to be computed such that for all
required initial states results are obtained for collision energies
sufficiently close to the predefined ``target collision energy'' grid given
by Eq.~(\ref{eq:Ecoll}) in Sec.~\ref{sec:Egrid}. A collision energy
is ``sufficiently close'' if it deviates at most by a fraction $s=0.1$,
see Eq.~(\ref{eq:DeltaE}).

The algorithm works like this. First, we sort the channel energies such that
\begin{equation}
\epsilon_1 \leq \epsilon_2 \leq \ldots \leq \epsilon_n.
\end{equation}
Next, we start with the highest ``target collision energy'' $E^\mathrm{coll}_k$
with $k=k_\mathrm{max}$ and
divide the energy interval $[\epsilon_1,\epsilon_n]$ into $N$ 
sub-intervals of equal width, with $N$ the smallest integer such that
\begin{equation}
  \frac{\epsilon_n-\epsilon_1}{N} \leq s E_k^\mathrm{coll},\;\;\;
  \mbox{for $k=k_\mathrm{max}$}.
\end{equation}
If there is only a single channel energy $\epsilon_i$ in a given sub-interval,
then we do a separate CC calculation for each
\begin{equation}
  E^\mathrm{tot}_k = \epsilon_i+E_k^\mathrm{coll},\;\;\;
  \mbox{for $k=1,2,\ldots k_\mathrm{max}$}
\end{equation}
and we use these calculations only for the channel with energy $\epsilon_i$.
For sub-intervals containing multiple channel energies
\begin{equation}
\{\epsilon_l,\epsilon_{l+1}, \ldots,\epsilon_m\}
\end{equation}
we compute a total energy as
\begin{equation}
  E^\mathrm{tot} = \frac{\epsilon_l+\epsilon_m}{2}
  +\sqrt{\left(\frac{\epsilon_m-\epsilon_l}{2}\right)^2
  +\left(E_k^\mathrm{coll}\right)^2}
\end{equation}
In case $\epsilon_l=\epsilon_m$ this gives
\begin{equation}
  E^\mathrm{tot} = \epsilon_l + E_k^\mathrm{coll}
\end{equation}
and otherwise it makes the geometric mean of the
collision energies corresponding the channels
$l$ and $m$ equal to the target collision energy
\begin{equation}
 E_k^\mathrm{coll} = \sqrt{ (E^\mathrm{tot}-\epsilon_l) (E^\mathrm{tot}-\epsilon_m)}.
\end{equation}
The total energies required for the next lower
target energy $E_{k-1}^\mathrm{coll}$ are found by applying
this algorithm recursively for each sub-interval $[\epsilon_l,\epsilon_m]$.
Finally, for the lowest collision energy $E_1^\mathrm{coll}$
we do a separate CC calculation for each channel energy.
This way, we avoid having collision energies lower than
the lowest target collision energy. In this low energy
region there may be resonances, for which a different
algorithm may be used.

\section{Modified renormalized Numerov algorithm}
\label{sec:Numerov}

The renormalized Numerov algorithm was presented by Johnson \cite{johnson:79}
and a slightly different implementation was
given by Hubbard {\em et al.} in the appendix of Ref.~\onlinecite{hubbard:83}.
Here we derive a variation that uses less memory than the
Hubbard implementation, and less cpu time per propagation step
than the Johnson version.

The CC equations, Eqs.~(A1) and (A4) from Hubbard {\em et al.}, in
our notation, are
\begin{align}
    &\frac{d^2}{dR^2}\bmc(R)  =\bmW(R)\bmc(R)\\
    &\bmW_n  = \bmW(R_n)   =    \frac{2\mu}{\hbar^2}[\bmV(R_n)-E \bmI)],
\end{align}
where $\bmc(R)$ is a vector with expansion coefficients of
the scattering wave function in the BF basis 
given in Sec.~\ref{sec:hamiltonian}, $\bmV(R_n)$ is the potential
energy matrix, and $\bmI$ is the identity matrix.
With
\begin{equation}
    \label{eq:fn}
    \bmF_n\equiv\bmI-\frac{h^2}{12}\bmW_n
\end{equation}
Eq.\ (A2) of Hubbard {\em et al.} becomes
\begin{equation}
    \bmF_{n-1}\bmc_{n-1}-(12\bmI-10\bmF_n)\bmc_n+\bmF_{n+1}\bmc_{n+1}={\bm 0}.
\end{equation}
Equations (26) and (27) of Johnson\cite{johnson:79}, in our notation, are
\begin{align}
    \tilde{\bmc}_n&=\bmF_n\bmc_n\,, \\
    \tilde{\bmc}_{n-1}-\bmU_n&\tilde{\bmc}_n+\tilde{\bmc}_{n+1}=0\,,
\end{align}
where $\tilde{\bmc}_n$ corresponds to what Johnson \cite{johnson:79} denotes as $\bmF_n$, and from their Eq. (28), we obtain:
\begin{align}
    \bmU_n=\bmF_n^{-1}(12\bmI-10&\bmF_n)=12\bmF_n^{-1}-10\bmI,\\
    \widetilde{\bmQ}_n=\tilde{\bmc}_{n-1}\tilde{\bmc}_n^{-1}&=\bmF_{n-1}\bmQ_n\bmF_n^{-1}\,,
\end{align}
where we use Eq.~(A6) of Hubbard {\em et al.} to define:
\begin{equation}
    \bmQ_n=\bmc_{n-1}\bmc_n^{-1}\,.
\end{equation}
Johnson's\cite{johnson:79} recursion relation becomes:
\begin{equation}
    \widetilde{\bmQ}_n-\bmU_n+\widetilde{\bmQ}_{n+1}^{-1}={\bm 0},
\end{equation}
which can be rearranged to:
\begin{equation}
    \widetilde{\bmQ}_{n+1}=(\bmU_n-\widetilde{\bmQ}_n)^{-1},
\end{equation}
where Johnson\cite{johnson:79} denotes $\bmQ_n$ as $\bmR_{n-1}^{-1}$ in his Eq.~(30).
The iteration is started with
\begin{equation}
    \widetilde{\bmQ}_1={\bm 0}.
\end{equation}
For the iterations $n = 1, 2,\ldots,N-2$, we have:
\begin{equation}
  \label{eq:Qn1}
    \widetilde{\bmQ}_{n+1}=\left[12\bmF_n^{-1}-10\bmI-\widetilde{\bmQ}_n\right]^{-1}.
\end{equation}
In the final step ($n=N$), we compute:
\begin{equation}
    \bmQ_n=\bmF_{n-1}^{-1}\widetilde{\bmQ}_n\bmF_n\,,
\end{equation}
which can be written as:
\begin{equation}
    \bmQ_n=\left[12\bmI-(10\bmI+\widetilde{\bmQ}_{n-1})\bmF_{n-1}\right]^{-1}\bmF_n.
\end{equation}
Recursion relation Eq.~(\ref{eq:Qn1}) requires two matrix inversions.
We rewrite it as
\begin{equation}
\label{eq:recur}
    \widetilde{\bmQ}_{n+1}=\left[12\bmI-\bmF_n(10\bmI+\widetilde{\bmQ}_n)\right]^{-1}\bmF_n,
\end{equation}
so that each iteration requires a matrix multiplication and solving
a set of linear equations, which is faster than two matrix
inversions. Note that the potential matrix
$\bmV(R_n)$ is symmetric, and hence also the matrices $\bmW_n$, $\bmU_n$,
and $\widetilde{\bmQ}_n$. This can be exploited to speed up solving
the linear equations.

When we compare our final recursion relation Eq.~(\ref{eq:recur}) with
Eq.~(A6) of Hubbard {\em et al.} we find that we do not have to
keep matrices $\bmF_{m-1}$ and $\bmF_{m-2}$  from previous iterations.
Comparing our expression to Eqs.~(28) and (30) of Johnson, we find
that we have replaced one matrix inversion by a matrix multiplication.

Computing the cross sections involves transforming the $\bmK$-matrix to
the $\bmS$-matrix [Eq.~(21) in Ref.~\onlinecite{selim:22}].
We may rewrite that expression as a set of linear equations,
\begin{equation}
  (\bmI - i \bmK) \bmS = \bmI + i \bmK.
\end{equation}
Only solving these
equations for open channels and for the columns of the $\bmS$-matrix
that are needed reduces the memory requirements, in particular
since the $\bmS$-matrix is complex valued.
\bibliography{vanderwaals}
\end{document}